\documentclass[journal]{IEEEtran}

\usepackage{cite}

\usepackage{dsfont}

\ifCLASSINFOpdf
   \usepackage[pdftex]{graphicx}
\else

\fi

\usepackage{bm}
\usepackage{amsmath,amssymb}

\usepackage{algorithm}
\usepackage[exponent-product = \cdot]{siunitx}

\usepackage{url}
\usepackage{hyperref}

\newcommand{\sysX}{\bm x}
\newcommand{\sysU}{\bm u}
\newcommand{\sysY}{\bm y}
\newcommand{\pathParam}{s}
\usepackage[mode=buildnew]{standalone}

\newcommand{\objSet}{\mathcal{J}}
\newcommand{\paramSet}{\Theta}
\newcommand{\paramSetFeasible}{\paramSet_{\mathrm{f}}}

\newcommand{\params}{\boldsymbol{\theta}}

\DeclareMathOperator*{\argmax}{arg\,max}

\usepackage{color}
 
\begin{document}

\title{What is the Best Way to Optimally Parameterize the MPC Cost Function for Vehicle Guidance?}

\author{David Stenger\textsuperscript{1}, Robert Ritschel\textsuperscript{2}, Felix Krabbes\textsuperscript{3}, Rick Voßwinkel\textsuperscript{3}, and Hendrik Richter\textsuperscript{4}
\thanks{\textsuperscript{1}Institute of Automatic Control (IRT), RWTH Aachen University, Germany, {\tt d.stenger@irt.rwth-aachen.de}}
\thanks{\textsuperscript{2}Department Automated Driving Functions, IAV GmbH, Germany, 
{\tt robert.ritschel@iav.de}}
\thanks{\textsuperscript{3}Faculty of Automotive Engineering, Zwickau University of Applied Sciences, Germany,   {\tt felix.krabbes.lzc@fh-zwickau.de, rick.vosswinkel@fh-zwickau.de}}
\thanks{\textsuperscript{4}Faculty of Engineering, HTWK Leipzig University of Applied Sciences, Germany,   {\tt hendrik.richter@htwk-leipzig.de}}}

\maketitle

\begin{abstract}
Model predictive control (MPC) is a promising approach for the lateral and longitudinal control of autonomous vehicles. However, the parameterization of the MPC with respect to high-level requirements such as passenger comfort as well as lateral and longitudinal tracking is a challenging task. Numerous tuning parameters as well as conflicting requirements need to be considered. This contribution 
formulates the MPC tuning task as a multi-objective optimization problem. Solving it is challenging for two reasons: First, MPC-parameterizations are evaluated on an computationally expensive simulation environment. As a result, the used optimization algorithm needs to be as sample-efficient as possible. Second, for some poor parameterizations the simulation cannot be completed and therefore useful objective function values are not available (learning with crash constraints). In this contribution, we compare the sample efficiency of multi-objective particle swarm optimization (MOPSO), a genetic algorithm (NSGA-II) and multiple versions of Bayesian optimization (BO). We extend BO, by introducing an adaptive batch size to limit the computational overhead and by a method on how to deal with crash constraints. Results show, that BO works best for a small budget, NSGA-II is best for medium budgets and for large budgets none of the evaluated optimizers is superior to random search. Both proposed BO extensions are shown to be beneficial. 
\end{abstract}


\IEEEpeerreviewmaketitle
\section{Introduction}

In the last few decades, automated and autonomous driving became an important topic in automotive technology. Particularly good results have been obtained in limited operational design domains,  
such as highway driving and park scenarios.  
In this context, numerous approaches to longitudinal and lateral control have emerged for driver assistance systems and automated driving functions (e.g. \cite{ulsoy_peng_cakmakci_2012, kiencke2005automotive,isermann2022automotive, No2000, Liaw2008, Zhang2009}). 

As it became desirable to extend automation and assistance functions into unstructured situations and more complex traffic scenarios, optimization-based approaches, especially model predictive control (MPC), have been established for vehicle guidance~\cite{Hrovat2012}.
MPC allows us to define the resulting behavior of the vehicles using a cost function and a prediction horizon. 
However, the manual tuning of the MPC cost function is challenging for complex high-level criteria such as comfort.

As an alternative, numerous optimization approaches have been presented for single-objective MPC tuning (e.g. \cite{YAMASHITA2016, Shah2011, Vega2007, ALGHAZZAWI2001, stenger2020robust, Ramasamy2019}).
Instead this paper gives one example on how to formulate the tuning task as a multi-objective black-box optimization problem. 
Multi-objective optimization allows us to simultaneously address conflicting objectives such as comfort and tracking accuracy. In order to asses an MPC parameterization on a meaningful driving cycle, an expensive-to-evaluate closed-loop simulation of the vehicle in its environment is used. As an additional challenge, objective function values are not available for some parameterizations due to failing simulations (crash constraints \cite{Marco.2021}). Because the simulation environment is a black-box to the optimizer, objectives, tuning parameters, and models can be easily exchanged with the presented approach.     

For single-objective optimization and limited budgets, Bayesian optimization (BO) has been shown to require less expensive simulations (i.e. it has a better sample-efficiency) than population-based metaheuristics for multiple control related applications \cite{JournalPreprint}. 
Therefore, multi-objective BO (MOBO) is a promising candidate for the multi-objective tuning problem. MOBO has seen a fair amount of applications in the control and robotics community (e.g. \cite{AriizumiRyo.2017, TeschMatthew.2013, TurchettaMatteo.2020, YeonjuKim.2021}). In our previous work \cite{GharibAli.2021}, we also introduced MOBO for MPC parameter tuning.  

However, the choice of acquisition function heavily influences the sample-efficiency and computational overhead of MOBO. Most of the previous work on MOBO for controller tuning \cite{AriizumiRyo.2017, TeschMatthew.2013, TurchettaMatteo.2020} use the so called expected improvement of hypervolume (EIHV) acquisition function \cite{M.Emmerich.2008}. However, the expected improvement-matrix (EIM) criterion \cite{Zhan.2017} as well as Thompson sampling efficient multi-objective optimization (TSEMO) \cite{Bradford.2018} were shown to have a similar sample-efficiency with less computation overhead than EIHV \cite{Zhan.2017, Bradford.2018}. In previous work with EIM \cite{GharibAli.2021}, the computational overhead quickly grew with an increasing amount of objective function evaluations. In contrast to EIM and EIHV, TSEMO comes with a straightforward heuristic for batch evaluations \cite{Bradford.2018} and is therefore expected to scale better.      

In previous works on MOBO for controller tuning, it has not been studied which version of BO is more sample-efficient than metaheuristics. 
Therefore, our contribution compares the sample-efficiency and the computational overhead of five different MOBO versions and two metaheuristics, Multiple Objective Particle Swarm Optimization (MOPSO) and Non-dominated Sorting Genetic Algorithm II (NSGA-II) on the vehicle guidance task. Additional benchmarks are random search and grid search. Furthermore, we extend TSEMO to address crashing simulations and adaptive batch sizes.

The paper is structured as follows. After the introduction we give a more detailed definition of the problem in Section~\ref{sec:problem}. Afterwards, Section \ref{sec:MPCTuning} describes and discusses the MPC tuning framework used. The investigated algorithms are presented in Section~\ref{sec:algorithms}, which is followed by benchmark results in Section ~\ref{sec:Results}. Results also highlight the practical applications of the presented approach. After that, the paper concludes with a summary of the results (Section \ref{sec:conclusion}).

\section{Problem Statement \label{sec:problem}}

In order to address the MPC tuning problem, we formulate it as a deterministic multi-objective optimization problem 
\begin{equation}
	\begin{aligned} \label{eq:ProbMultOpt} 
  \  \min_{\params \in \mathbb{R}^N} \bm J(\params) = &  \left[ J_1(\params), J_2(\params), \dots , J_M(\params)  \right] \\
		\mathrm{s.t. } \qquad \qquad \qquad  & \params_{\mathrm{min}} \le \params \le \params_{\mathrm{max}} \\
		& l(\params) = 1  \mathrm{.} \\ 
	\end{aligned}
\end{equation}
with $M\geq2$ individual objective functions and a vector of $N$ decision variables $\params = \left[\theta_1, \theta_2, \dots , \theta_N \right]$.
The individual objective functions $J_i(\params): \mathbb{R}^N \rightarrow \mathbb{R}$, $i\in \{1,2, \dots M\}$ describe several aims or targets of the overall control behavior and the decision variables $\theta_i$ are specific MPC parameters. The vectors of the upper and lower bounds of the decision variables are denoted with $\params_{\mathrm{max}}$ and $\params_{\mathrm{min}}$, respectively.
The set 
\begin{equation}
   \paramSetFeasible = \left\{\params \in \mathbb{R}^N \mid \params_{\mathrm{min}} \le \params \le \params_{\mathrm{max}};\, l(\params) = 1 \right\}
\end{equation}
is the feasible set of decision vectors and represents the feasible domain in the decision space.

In addition to the objective function values, it is also crucial for the valuation of a MPC parameterization whether it causes the system to crash or not. To formalize this criterion we define the function $l(\params)$, which returns 1 in the positive case and 0 in the crash case. The function is part of an equality constraint of the optimization problem \eqref{eq:ProbMultOpt}, termed crash constraint, and thus has direct influence on the feasible domain $\paramSetFeasible$. In case of a system crash $l(\params) = 0$,  objective function values are not available. Note that in general there is no analytical description of the function $l(\params)$ for the MPC tuning problem. This setting is also known as learning with crash constraint \cite{Marco.2021}.   

In multi-objective optimization, there is usually no feasible solution that minimizes all objective functions simultaneously. Generally, if one objective is improved, other objectives are degraded. So, the superiority of a solution over other solutions can not be determined by just comparing their objective function values, like in the single-objective optimization case. Therefore, the focus is on Pareto optimal solutions. Here, the goodness of a solution is determined by the dominance. A feasible solution $\params_a\in \paramSetFeasible$ is said to dominate another solution $\params_b\in \paramSetFeasible$, denoted with $\params_a\prec \params_b$, if and only if
\begin{align}
   \forall i \in \{1, 2, \dots, M\}&: J_i(\params_a) \leq J_i(\params_b) \text{ and}\notag\\
   \exists j \in \{1, 2, \dots, M\}&: J_j(\params_a) < J_j(\params_b).
\end{align}
A solution $\params^* \in \paramSetFeasible$ is said Pareto optimal if there does not exist another solution in the feasible domain that dominates it. The set of all Pareto optimal solutions in the feasible domain, denoted $\Theta_{\mathrm{po}}$, is called the Pareto optimal solution set and is defined as:
\begin{equation}
   \Theta_{\mathrm{po}} = \left\{\params^* \in \paramSetFeasible \mid \neg\exists\,\params \in \paramSetFeasible: \params \prec \params^* \right\}.
\end{equation}
The set of objective function vectors corresponding to the Pareto optimal solution set $\Theta_{\mathrm{po}}$ is the Pareto front and is defined as:
\begin{equation}
   \label{eq:ParetoFront}
   \objSet_{\mathrm{po}} = \left\{ \bm J(\params^*) = \left[ J_1(\params^*), \dots , J_M(\params^*)  \right] \mid \params^* \in \Theta_{\mathrm{po}} \right\}.
\end{equation}
This set is also known as the Pareto optimal frontier or Pareto equilibrium surface.

Thus, the result of solving the multi-objective optimization problem \eqref{eq:ProbMultOpt} is the Pareto optimal solution set $\objSet_{\mathrm{po}}$, which is a set of solutions that define the best tradeoff between competing objectives. The knowledge of the Pareto optimal solution set and its corresponding Pareto front $\objSet_{\mathrm{po}}$ allows to select specific MPC parameterizations for specific use cases afterwards. For this purpose, it is necessary to decide which objectives are of particular importance for a specific use case. 

In general, solving such a multi-objective optimization problem is very time-consuming and provides only an approximation of the Pareto front. Therefore, it is important to use suitable approaches for this purpose, which provide a good approximation of the Pareto front with low computational effort. The comparison of different approaches for MPC tuning is the main focus of this paper.

In this contribution, one MPC parameterization $\params$ is evaluated by using a closed-loop simulation of the vehicle in its environment. This poses two challenges for the multi objective optimization algorithm. First, an analytical description of the objective function is not available. Instead the optimizer can only query the simulation environment with parameters and record the responses. This setting is known as black-box optimization. Second, each simulation takes a considerable amount of time. Therefore the sample-efficiency i.e. the ability of an optimizer to approximately solve \eqref{eq:ProbMultOpt} with as little objective function evaluations as possible is of major importance.      

\section{Simulation-based MPC Tuning for Vehicle Guidance} \label{sec:MPCTuning}
This section will introduce the specific MPC tuning use case studied in this paper. The task is to optimize the parameterization of a model predictive controller for the longitudinal and lateral guidance of vehicles using a simulation.

\subsection{Controller Tuning Framework}
\begin{figure}
	\centering
	\includegraphics[width=1.0\linewidth]{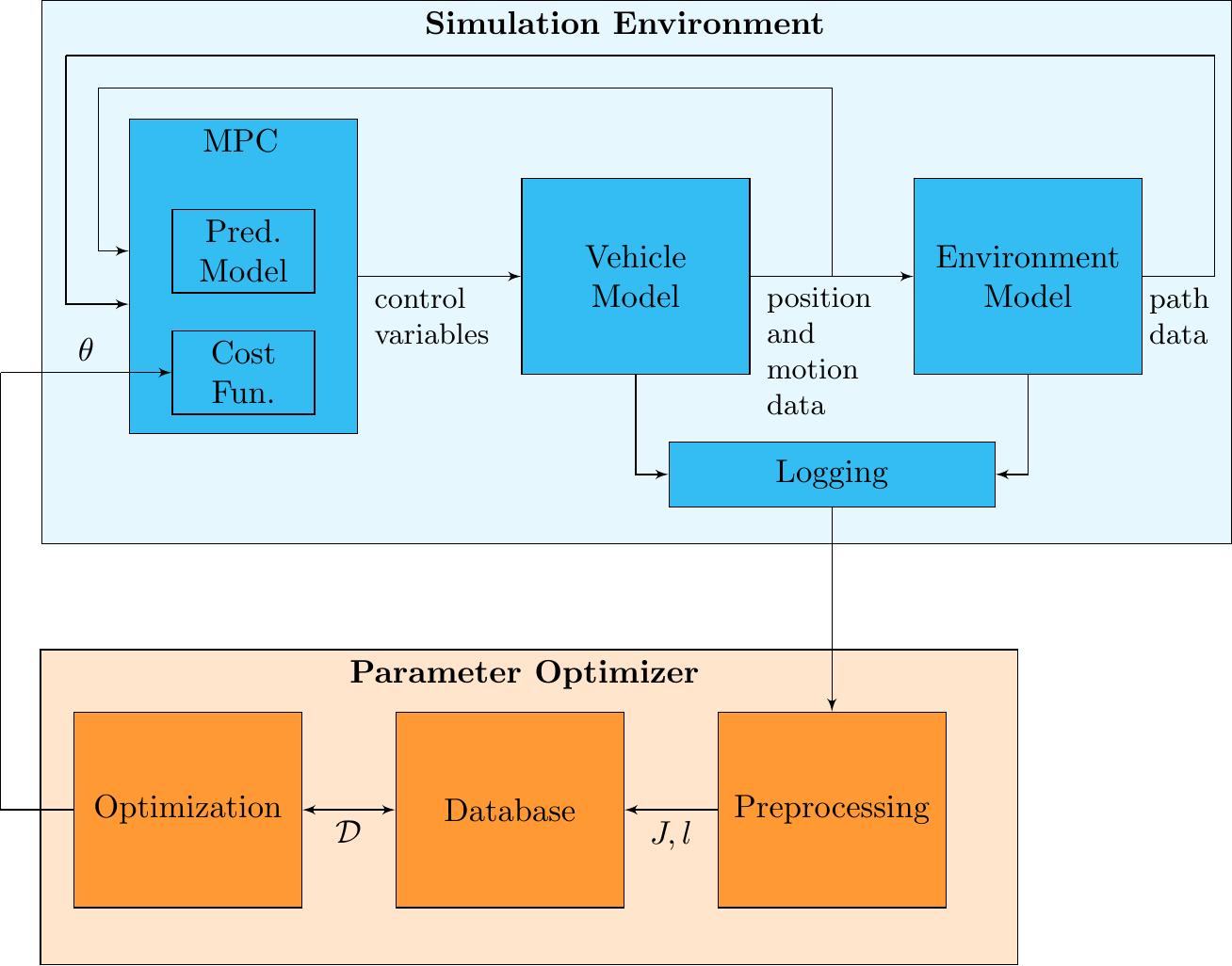}
	\caption{Controller Tuning Framework}
	\label{fig:ControllerTuningFramework}
\end{figure}

In order to find optimal MPC parameterizations we propose a controller tuning framework as shown in Fig.~\ref{fig:ControllerTuningFramework}, which consists of tow main parts: the parameter optimizer and the simulation environment. Please note that we use the framework for MPC parameter tuning, but it can easily be adapted for parameter tuning of any other type of controller. 

The parameter optimizer handles the complete optimization process. This includes 1) receiving, storing and pre-processing the data of the simulation (i.e. calculating the objective function values from the time domain data), 2) invoking the optimization algorithm with the current data as well as 3) querying the simulation environment with a specific parameter vector $\params$. How often and in which order these three actions are performed depends on the multi-objective optimization algorithm used.

The simulation environment is utilized to evaluate control performance of the MPC for a given parameter vector $\params$. For this purpose the MPC is simulated in closed-loop with a nonlinear plant. The plant consists of two main parts: a vehicle model and an environment model. The vehicle model is a nonlinear dynamic model and describes the lateral and longitudinal behavior of the ego vehicle in dependence of the control inputs and the feedback from the environment model. Its outputs are the states of the ego vehicle, such as the velocity, the position or the orientation. Based on the states of the ego vehicle, the environment model provides input data for the MPC controller. In our case, this is lane data obtained using idealized lane detection for a predefined virtual road with two lanes. Furthermore, for receiving the input data and providing the simulation data, the simulation environment has appropriate interfaces to the parameter optimizer.

The simulation environment is implemented using the tool MATLAB/Simulink. As mentioned above, the duration of the simulation has a crucial impact on the overall duration of the tuning. Therefore, we minimize the simulation duration by exporting the Simulink model as an executable that runs fast, standalone simulations. This is done using the rapid simulation (RSim) feature of Simulink and allows repeat simulations with varying inputs without rebuilding the model. 

\subsection{MPC for Vehicle Guidance}
The subject of our studies is the parameter tuning of the MPC realization presented in \cite{Ritschel2019}. This MPC realization is designed for longitudinal and lateral vehicle control and is based on the so-called model predictive path-following control (MPFC) approach, a nonlinear MPC approach presented in \cite{Faulwasser.2013}. We will provide a brief overview of this MPC realization in the following. For a more comprehensive description, we refer to \cite{Ritschel2019}.

The main idea of this MPC realization is that the system, i.e. the ego vehicle, follows a given geometric reference path that specifies the desired position (e.g. the coordinates of the driving lane center) and the desired orientation of the vehicle. The reference path is given as a parametrized regular curve $\bm p: \pathParam \in [0,\pathParam_{\mathrm{max}}] \mapsto \mathbb{R}^{3}$, where $\pathParam \in \mathbb{R}$ is the the so-called path parameter. The behavior of the ego vehicle is modeled as a continuous-time nonlinear system with the control input $\sysU\subseteq\mathbb{R}^{2}$, the output vector $\sysY\in\mathbb{R}^{3}$ and the state vector $\sysX\subseteq\mathbb{R}^{9}$. 

The task of the MPC controller is to simultaneously determine the control input $\sysU(t)$ as well as the time evolution of path parameter $\pathParam(t)$ in such a way that the ego vehicle follows the reference path as closely as possible. To solve this path-following problem a sampled-data MPC strategy is used.

The control input $\sysU(t)$ is obtained via the repetitive solution of an optimal control problem in a receding horizon fashion. The cost functional $C$ to be minimized at every discrete sampling time instance $t_k = k\,T_s$ is given by
\begin{align}
	\label{eq:MPCCostFunction}
	C(\sysX(t_k),\,\pathParam(t_k),\bar{\sysU}(\cdot),&\bar{v}_s(\cdot))
	=\\
	\int\limits_{t_k}^{t_k+T_p}\! \left\|
		\begin{matrix}
	\bar{\bm e}(\tau)\\
	a_{\mathrm{lat}}(\bar{\sysX}(\tau))\\
	\end{matrix}\right\|^2_Q &+  \left\|
		\begin{matrix}
	\bar{\sysU}(\tau) \\
	\bar{v}_{\mathrm{s}}(\tau) - v_{\mathrm{s,des}}(s(\tau))\\
	\end{matrix}\right\|^2_R	\, \mathrm{d}\tau \notag \\
		&+ \left\|
			\begin{matrix}
	\bar{\bm e}(t_k+T_p)\\
	a_{\mathrm{lat}}(\bar{\sysX}(t_k+T_p))\\
	\end{matrix}\right\|^2_P \notag
\end{align} 
where the predicted system states and inputs are denoted by $\bar{\sysX}(\cdot)$ and $\bar{\sysU}(\cdot)$, respectively. In addition, $T_p$ is the prediction horizon, $\bm e(t) = \bm \sysY(t) - \bm p(\pathParam(t))$ is the deviation from the path, $a_{\mathrm{lat}}(\sysX)$ is an approximation of the lateral acceleration of the ego vehicle and $v_{\mathrm{s}}(t) = \dot{\pathParam}(t)$ is the path velocity. Consequently, the controller minimizes the path deviation and lateral acceleration while trying to achieve a given path velocity $v_{\mathrm{s,des}}$.

As common in model predictive control, $Q$ and $R$ indicate the stage cost weighting matrices and $P$ is the terminal cost weighting matrix. The weighting matrices are chosen as
\begin{equation}
\begin{array}{l}
Q = \operatorname{diag}(q_x, q_y, q_\psi, q_a), 
\\
P = \operatorname{diag}(p_x, p_y, p_\psi, p_a), 
\\    
R= \operatorname{diag}(r_a, r_\omega, r_v),
\end{array}
\label{eq:MPC_Q_R_P}
\end{equation}
where $q_x, q_y, q_\psi$ and $q_a$ are the weighting factors for x-, y-, yaw deviations and the lateral acceleration, respectively. The factors $p_x, p_y, p_\psi$ and $p_a$ do represent the same as the latter but for the terminal cost. Furthermore, $r_a, r_\omega$ and $r_v$ respectively indicate the weighting factors for the control input $\sysU(t)$, i.e., the target longitudinal acceleration, the steering wheel target angular velocity, and the path velocity error, respectively.

\subsection{Multi-Objective Optimization Problem Formulation}
In this paper, we focus on the tuning of the weighting factors \eqref{eq:MPC_Q_R_P} of the MPC cost functional \eqref{eq:MPCCostFunction}. The cost functional has eleven different weighting factors. If each factor is considered as a separate decision variable, this would lead to a complex optimization problem that is time-consuming to solve. Therefore, we perform various simplifications to reduce the solution time of the problem.
First, the weights in $Q$ and $P$ are set equal in order to reduce the number of decision variables. Second, we weight the deviation from the path in the x- and y-directions equally, which means $q_x = q_y$. Finally, we set the weighting factor $q_\psi$ to a fixed value, since the absolute value of the weighting factors is not important for the cost function, but only the relation of the factors to each other. Furthermore, we do not directly treat the weighting factors as decision variables, but we optimize the exponent of an exponential term, because we expect that changing a weighting factor, e.g., from 1 to 10 will have a similarly high impact as changing it from 10 to 100. Thus, we get $N=5$ decision variables for which
\begin{align}
 q_x = q_y = p_x = p_y &= 10^{\theta_1}  & r_\omega &= 10^{\theta_4} \notag\\
 q_a = p_a &= 10^{\theta_2} & r_\vartheta &= 10^{\theta_5}\\
 r_a &= 10^{\theta_3} \notag
\end{align}
holds. As boundaries for the decision variables, we use
\begin{align} 
   \params_{\mathrm{min}} &= \left[\theta_{\mathrm{min},1}, \dots , \theta_{\mathrm{min},5} \right] = \left[-3, \dots , -3 \right],\notag\\
    \params_{\mathrm{max}} &= \left[\theta_{\mathrm{max},1}, \dots , \theta_{\mathrm{max},5} \right] = \left[4, \dots , 4 \right].
\end{align}

In order to achieve the desired control behavior, appropriate objective functions must be defined. For this purpose, we focus on the driving behavior in this work. Different situations, such as urban driving, highway driving or parking, require different control behavior to ensure the desired driving quality. On the one hand we consider the tracking behavior of the lateral and longitudinal setpoints. On the other hand we concentrate on comfort, since vehicle guidance needs a very well applied control that passengers accept the automation.
Note that we use only three objective functions for this paper because the selected objectives are reasonable for tuning a vehicle guidance MPC based on practical experience. In general, any number of objective functions can be used.

\begin{figure}
	\centering
	\includegraphics[width=1.0\linewidth]{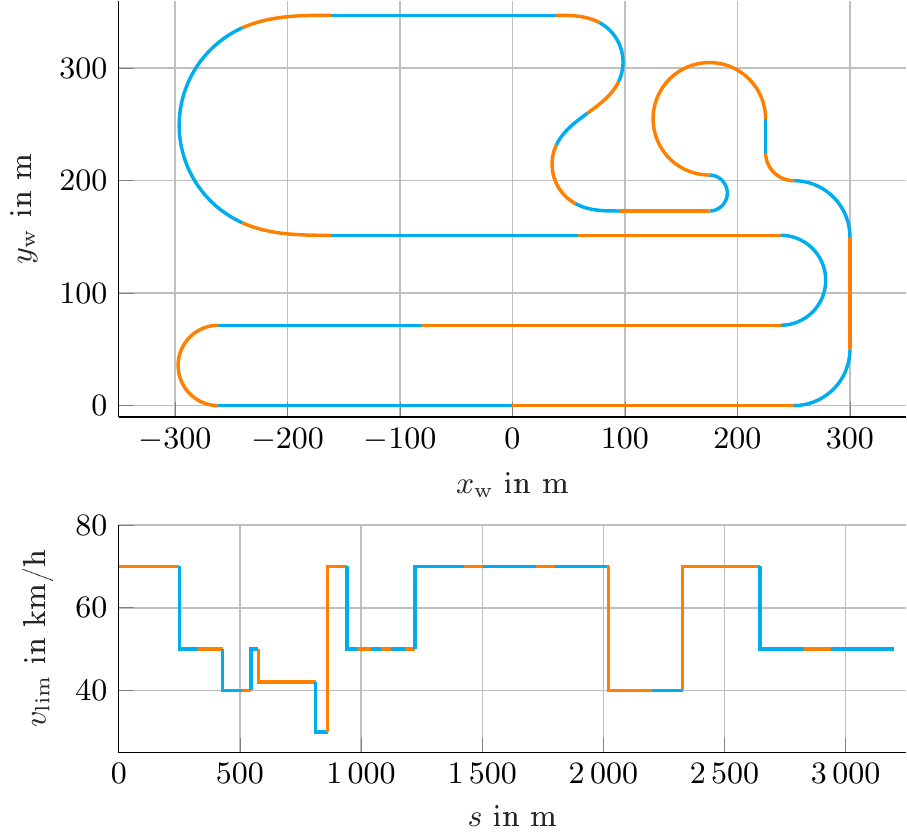}
	\caption{2D view and speed limit $v_{lim}$ of the road used in the simulation. The different sections of the road are highlighted by the colors cyan and orange for clarity.} 
	\label{fig:BridviewOfTheTrack}
\end{figure}

To quantify the driving behavior, we use the outputs of the simulation. For the following studies of the different optimization algorithms, we simulate an urban-like scenario that represents driving on the two-way road shown in Fig.~\ref{fig:BridviewOfTheTrack}, equivalent to the publication \cite{GharibAli.2021}. We assume that each call of the simulation with the parameterization $\params$ returns trajectories of length $N_k \in \mathbb{N}$, where $N_k$ can be different for each parameterization $\params$. We describe the trajectories as finite sequences. For example, let $\left(a(t_k;\params)\right)_{k=1}^{N_k}$ be the trajectory of the acceleration, where $a(t_k;\params)$ is the element for the sampling time instance $t_k = k\,T_s$ using parameterization $\params$.

\emph{Longitudinal Tracking:}
To describe the longitudinal tracking behavior we use the objective function
\begin{align} 
   J_{1}(\params) &=  \sqrt{\frac{1}{N_k}\sum_{k=1}^{N_k}(v_{\mathrm{des}}(t_k;\params)-v(t_k;\params))^2},
   \label{eq:errorV}
\end{align}
which is the root-mean-square error of the deviation between the ego velocity and the desired velocity. It is calculated using the returned trajectories of the desired and driven ego velocity $\left(v_{\mathrm{des}}(t_k;\params)\right)_{k=1}^{N_k}$ and $\left(v(t_k;\params)\right)_{k=1}^{N_k}$, respectively.

\emph{Lateral Tracking:}
The lateral tracking performance is quantified with the root-mean-square error of the lateral deviation from the desired path $e_{\mathrm{lat}}$ using the trajectory $\left(e_{\mathrm{lat}}(t_k;\params)\right)_{k=1}^{N_k}$ returned by the simulation. Thus, the second objective function is
\begin{align} 
    J_{2}(\params) =& \sqrt{\frac{1}{N_k}\sum_{k=1}^{N_k}(e_{\mathrm{lat}}(t_k;\params))^2}.
	\label{eq:errorLat}
\end{align}

\emph{Comfort:} The study \cite{Wang2020-dd} shows that vehicle acceleration has a very high impact on passenger comfort. It was found that large amplitudes of lateral and longitudinal acceleration of the vehicle caused passengers to feel discomfort. Therefore, we choose 
\begin{align} 
    J_{3}(\params) = \sqrt{\frac{1}{N_k}\sum_{k=1}^{N_k}(a(t_k;\params))^2}
   	\label{eq:errorJerk}
\end{align}
as the third optimization objective in order to minimize the vehicle acceleration $a=\sqrt{a_\mathrm{lat}^2+a_\mathrm{long}^2}$. The acceleration trajectory of the ego vehicle returned by the simulation is denoted with $\left(a(t_k;\params)\right)_{k=1}^{N_k}$. 

Finally, as function to describe whether a MPC parameterization causes the system to crash or not, we use
\begin{align} 
    &l(\params) =  
    \begin{cases}
    1 & e_{\mathrm{lat,max}}(\params) \leq 1.5 \wedge N_\mathrm{laps}(\params) = 1 \\
    0 & \text{otherwise}
    \end{cases}\\
    \intertext{with}
    &e_{\mathrm{lat,max}}(\params) = \max_{k\in \{1 \dots N_k\}} e_{\mathrm{lat}}(t_k;\params).
\end{align}
Here, $N_\mathrm{laps}$ is the number of laps driven by the ego vehicle. Consequently, the control task is successfully solved if the vehicle completes a lap and does not exceed a maximum lateral deviation of $\SI{1.5}{\meter}$.

\section{Multi-Objective Optimization Algorithms \label{sec:algorithms}}

\subsection{Bayesian Optimization}

Algorithm \ref{Algo:BOMultObjCrashConstrnt} outlines the procedure of MOBO with crash constraints. In general it can be also used for for non-deterministic optimization problems although it is applied to a deterministic formulation here. For a general introduction to BO, the reader is referred to \cite{Shahriari.2016}. For a better readability we define the operator $\tilde\cup$ where $\tilde t= t_1\ \tilde\cup \ t_2$ is the tuple with the ordered elements of $t_1$ first followed by the ordered members of $t_2$, i.e. $(a,b,c)\ \tilde\cup \ (d,e) = (a,b,c,d,e)$ .
\begin{algorithm}[ht] 
	1: Generate initial data \hbox{$\mathcal{D}_1 = \left( \paramSet_1,  \objSet_{1,1},  \dots , \objSet_{1,M}, \mathcal{L}_{1} \right)$}  \\ [3pt]
	2: \textbf{for} k = 1, 2, \dots  \textbf{do} \\[3pt]
	3: \quad Calculate the set of non-dominated solutions $\objSet_{\mathrm{po},k}$\\[3pt] 
	4: \quad $(\hat{\mathcal{J}}_{k,1},  \dots , \hat{\mathcal{J}}_{k,M})    \leftarrow \textrm{addVirtualData}(\mathcal{D}_k, \objSet_{\mathrm{po},k} )$ \\[3pt] 
	5: \quad  Learn probabilistic surrogate models $(\mathcal{GP}_k^{\hat{J}_1}, \dots, \mathcal{GP}_k^{\hat{J}_M})$\\  
	 \hspace*{6.5mm} using training data  
	$(\mathcal{D}_k, \hat{\mathcal{J}}_{k,1},  \dots , \hat{\mathcal{J}}_{k,M})$ 
	\\[3pt]
	6: \quad $S_k  \leftarrow \textrm{calcBatchSize}$  \\[3pt]
	7: \quad Maximize acquisition function  
	\begin{equation*}
	\alpha_k \left(\Theta \right) = f \left(\Theta, \mathcal{GP}_k^{\hat{J}_1}, \dots,\mathcal{GP}_k^{\hat{J}_M}, \objSet_{\mathrm{po},k}\right)
	\end{equation*}
	\hspace*{6.5mm} to get $S_k$ new sample points: $\Theta_{k}^\prime = \argmax \alpha_k \left( \Theta \right)$. \\[3pt]	
	8: \quad Query  
	objective function with $\Theta_{k}^\prime = ( \boldsymbol{\theta}_{1}, \dots, \boldsymbol{\theta}_{S_k})$ \\ 
	 \hspace*{6.5mm} to obtain responses $(\objSet_{k,1}^\prime, \dots, \objSet_{k,M}^\prime,\mathcal{L}_{k}^\prime)$ \\[3pt]
	9: \quad Augment data with new evaluations: \\
 	\hspace*{6.5mm} \hbox{$\mathcal{D}_{k+1} = \left( \paramSet_k \ \tilde\cup \  \paramSet^\prime_k ,  \objSet_{k,1} \ \tilde\cup \  \objSet^\prime_{k,1},  \dots , \mathcal{L}_{k} \ \tilde\cup \  \mathcal{L}^\prime_{k}   \right)$}
 	 \\[3pt]    
	10: \textbf{end for} 
	\caption{Multi-objective Bayesian optimization with crash constraints and flexible batch size. \label{Algo:BOMultObjCrashConstrnt} }	
\end{algorithm}
In Step 1, an initial data set $\mathcal{D}_1$ is generated using $N_\mathrm{init}$ evaluations. It consists of the set of evaluated parameters $\paramSet_1 =(\params_{1,1},\dots, \params_{1,N_{\mathrm{init}}} )$ and the corresponding obtained objective function values $\objSet_{1,m}$ for each of the $m \in \{1, \dots M \}$ objectives. Information on which of the evaluations were successful and which crashed is stored in the set $\mathcal{L}_{1} = (l_1, \dots l_{N_\mathrm{init}})$, where $l_1$ corresponds to the first evaluated parameterization. In this contribution, $N_\mathrm{init} = 5$ initial samples are drawn at random. Afterwards, the main optimization loop is entered. At each iteration $k$, the current Pareto front $\objSet_{\mathrm{po},k}$ is calculated from past successful evaluations (Step 3). Virtual data points $\hat{\mathcal{J}}_{k,m}$ are calculated in Step 4 for all past crashed evaluations and objectives (cf. Sec. \ref{sec:VDP}).  

Afterwards,
virtual data points and successful evaluations are used  
to construct a fast-to-evaluate surrogate model $\mathcal{GP}_k^{\hat{J}_m}$ of each of the objective functions $J_m(\params)$ (Step 5). 
In this work, Gaussian Process Regression (GPR) \cite{Rasmussen.2006} is used as the surrogate model.
GPR is a non-parametric 
model which provides probabilistic predictions of the objective functions. 
In this work, models are defined by a constant mean function and an anisotropic squared exponential kernel. 
Hyperparameters are optimized at each iteration by maximizing the marginal log-likelihood.

Based on the GPR models and the current pareto optimal points, an acquisition function $\alpha_k \left( \Theta \right)$ 
is maximized 
to determine the $S_k$ next sample points $\Theta_{k}^\prime$ (Step 7). Here, two different acquisition functions are compared. TSEMO (cf. Sec. \ref{TSEMO}) can used with a variable batch size $S_k$. 
EIM (cf. Sec. \ref{EIM}) is used with a constant batch size of $S_k = 1$. After evaluating the new parameter combinations on the expensive-to-evaluate closed-loop simulation (Step 8), the data set is augmented (Step 9). This is done by adding the new parameter combinations $\Theta_{k}^\prime$ and the corresponding obtained objective function values $\objSet_{k,m}^\prime$ and crash constraint values $\mathcal{L}_{k}^\prime$ to the past evaluations $\mathcal{D}_k = \left( \paramSet_k,  \objSet_{k,1},  \dots , \objSet_{k,M}, \mathcal{L}_{k} \right)$ such that $\mathcal{D}_{k+1} = \left( \paramSet_k \ \tilde\cup \  \paramSet^\prime_k ,  \objSet_{k,1} \ \tilde\cup \  \objSet^\prime_{k,1},  \dots , \mathcal{L}_{k} \ \tilde\cup \  \mathcal{L}^\prime_{k}   \right)$. 

\subsubsection{Thompson Sampling Efficient Multi-Objective Optimization (TSEMO) with flexible batch size} \label{TSEMO}

In TSEMO  
\cite{Bradford.2018}, 
one sample from the GPR model $\mathcal{GP}_k^{\hat{J}_m}$ of each of the $M$ objectives is drawn. This way, an approximate objective function landscape, which is consistent to the observed data is generated at random. A multi-objective genetic algorithm, in this case \hbox{NSGA-II} \cite{Deb2002AFA}, searches for the pareto-optimal parameter set of the approximate objective function landscape. Because the sampled functions are fast-to-evaluate, this is orders of magnitudes cheaper than directly solving \hbox{Eq. \eqref{eq:ProbMultOpt}} with NSGA-II. From the Pareto-optimal candidate set, the parameterizations to be evaluated next on the expensive-to-evaluate black box are chosen by maximizing the increase in the hyper-volume indicator over $\objSet_{\mathrm{po},k}$. For our experiments, the implementation publicly available in \cite{Schweidtmann.2022} is used as a starting point. 

In \cite{Bradford.2018, Schweidtmann.2022}, the number of evaluations per iteration is constant. Instead, here at iteration $k$ the batch size $S_k$ is chosen adaptively to not exceed a desired relative overhead $p_{\mathrm{o,des}}$:
\begin{equation}
	S_k = \left\lceil \frac{t_{\mathrm{o},k-1}}{p_{\mathrm{o,des}} \ t_{\mathrm{sim},k-1}} \right\rceil \ ,
\end{equation}  
where $t_{\mathrm{o},k-1}$ is the overhead required for GPR training and determination of the next sample point during the previous iteration. The average time for one objective function evaluation during the last iteration is denoted as $t_{\mathrm{sim},k-1}$. Here $p_{\mathrm{o,des}} = 0.2$ is used.
While overhead is small, the ideal batch size of one is maintained to achieve optimal sample quality. With increasing overhead, a growing sample size trades more objective function evaluations for less ideal sample quality. We expect this trade-off to be beneficial.

\subsubsection{Expected Improvement Matrix Criterion} \label{EIM}

As an alternative to TSEMO, the EIM criterion presented in \cite{Zhan.2017} is used as a benchmark with a constant batch size of $S_k = 1$. It extends the well known single-objective expected improvement (EI) criterion to the multi-objective case. Here the euclidean distance-based variant $\mathrm{EIM_{e}}$ is employed. The $\mathrm{EIM_{e}}$ criterion was also used for multi-objective MPC tuning in previous work \cite{GharibAli.2021}.   

\subsubsection{Virtual data points (VDP) for multi-objective BO} \label{sec:VDP} In case of crashed simulations $(l(\params) = 0)$, the objective function values cannot be calculated from the simulation results.  
The most simple way to address this, is to assign constant objective function values.  
However, expert knowledge is required to decide which value to assign. Additionally, this intuitive approach leads to discontinuities at the borders between successful and unsuccessful evaluations, which is problematic, because GPR assumes smooth functions. 

Here, a heuristic based on pessimistic GP predictions, which has been used for the single-objective \cite{JournalPreprint} and constrained \cite{Stenger.2022} case is extended to the multi-objective case. MOBO with crash constraint was reported before in \cite{KentaKato.2017}, where an additional classifier was trained to predict crashes. A separate GP model was also used to predict crashed evaluations in \cite{GharibAli.2021}. In contrast, the approach presented here is compatible with any kind of acquisition function, because it only adapts the training data of the GPR models. Therefore, there is no need in the acquisition function to facilitate probability of feasibility as is the case in \cite {KentaKato.2017,GharibAli.2021}. 

The approach is summarized in Algo. \ref{Algo:CrashConstraint_MOBO}. First (Steps 1 and 2), GPR models for each objective are fitted using all successful evaluations (i.e. where $l_k = 1$ was observed). Afterwards, pessimistic GPR predictions are calculated for each of the crashed parameterizations (Step 4) using the predictive mean $\mu_{J_m}$ and standard deviation $\sigma_{J_m}$ of the model which was fitted with exclusively successful evaluations. Here the tuning parameter $\gamma$ is initialized to 3. Finally, the virtual data points are bounded to the worst observed successful evaluation for each objective $m$ (Step 5). If any of the virtual data points dominates the set of Pareto points $\objSet_{\mathrm{po},k}$, the tuning parameter $\gamma$ is increased (Steps 6 and 7). It is not desirable that the virtual data points are part of the Pareto front.

With this heuristic, the optimization is driven away from the region were crashes were observed. I.e. the expectation  at the virtual data points is worse than the current Pareto front, while uncertainty is reduced.  
At the same time, discontinuities are avoided, because smooth GP predictions are used to calculate the virtual data points.

\begin{algorithm}[t] 
	1: Extract all crashed evaluations: \\
		\hspace*{3.2mm} $\bar{\mathcal{D}}_k  = (\bar{\Theta}_k, \bar{\objSet}_{k,1}, \dots , \bar{\objSet}_{k,M}   )$   \\ [3pt]
		2: Fit GPR Models with successful evaluations 
		$\mathcal{D}_k \setminus \bar{\mathcal{D}}_k$. \\ [3pt]
	3: \textbf{for each} crashed query $\bar{\boldsymbol{\theta}} \in \bar{\Theta}_k$  \\ [3pt] 
	4: \quad Calculate virtual data point using a pessimistic \\  
	\quad \hspace*{6.5mm}  GP prediction:   $\hat{J}_m(\bar{\boldsymbol{\theta}}) = \mu_{J_m}(\bar{\boldsymbol{\theta}}) + \gamma \sigma_{J_m}(\bar{\boldsymbol{\theta}})$  \\ [3pt]
	5: \quad Bound pessimistic prediction to the worst successful \\     
	\quad \hspace*{6.5mm} evaluations
	$\hat{J}_m(\bar{\boldsymbol{\theta}}) = \mathrm{min} ( \hat{J}_m(\bar{\boldsymbol{\theta}}) , J_{\mathrm{max},m} ) $  \\ [3pt]
	6: \textbf{if} any virtual data points dominates one element in $\objSet_{\mathrm{po},k}$:\\  [3pt]
	7: \quad \textbf{do} $\gamma = \gamma + 0.5$; \textbf{Go to} line 3;     
	\caption{Calculation of virtual data points (Step 4 of \hbox{Algo. 1})}  
	\label{Algo:CrashConstraint_MOBO}
\end{algorithm}

\subsection{NSGA-II}
NSGA-II is a very popular evolutionary algorithm for multi-objective optimization \cite{branke.2008}. We based our implementation on the algorithm described in \cite{Deb2002AFA}. Algo. \ref{Algo:NSGAII} shows an outline of the procedure. After a initialization in Steps 1 to 3, the algorithm loops over the generation count $N_\mathrm{gen}$ (Step 4-12). In each iteration a new child generation is generated by mutation and crossover of the parent generation. In the crossover part, the parameters $\params$ of each individual in the child generation are set as a random point on the interpolation line between two members of the parent generation, shown in Step 5. After that in the mutation step, normal distributed random values are added to parameters from the crossover step in Step 6. After evaluating the new generation by the simulation procedure, the union of the child and the parent generation is sorted first by a non dominated sorting algorithm (Step 9) and then based on the crowding distance in Step 10. The last step is the decimation of the sorted entities to the given population count. 

To deal with the crashed simulations every cost value is set to $\infty$. By this, the non dominated sorting sets those parameter configurations at the lower end of the list to remove them in the truncation step where the following parent generation is generated from current parent and child generation which can be found in Step 11. If there are more invalid parameter configurations than elements that are deleted in this step, the invalid parameters are replaced by new random parameters for the next generation so mutation and crossover do not rely on parameter configurations that are known to be invalid.

The obtained results are based on a population size of $N_\mathrm{pop}=100$ and a total generation count of $N_\mathrm{gen}=50$.

\begin{algorithm}[t] 
	1: Generate an initial population $\Theta_{1} = (\boldsymbol{\theta}_{1,1}, \dots \boldsymbol{\theta}_{1,N_\text{pop}})$\\[3pt]
	2: Query objective function with $\Theta_{1}$ to obtain responses \\ 
       \hspace*{3.2mm} $(\objSet_{1,1}, \dots, \objSet_{1,M},\mathcal{L}_1)$  \\[3pt]
    3: Form initial Dataset $\mathcal{D}_{1}=(\Theta_{1}, \objSet_{1,1}, \dots, \objSet_{1,M},\mathcal{L}_1)$ \\[3pt]
	4: \textbf{for} each generation  $k = 1, 2, \dots ,N_\mathrm{gen}$ \textbf{do} \\[3pt]
	5: \quad Crossover: $\Theta_{k,\mathrm{Cross}}=\mathrm{Crossover}(\mathcal{D}_{k})$\\[3pt] 
	6: \quad Mutation: $\Theta_k^\prime=\mathrm{Mutation}(\Theta_{k,\mathrm{Cross}})$\\[3pt] 
	7: \quad Query objective function with $\Theta_{k}^\prime = (\boldsymbol{\theta}_{1}, \dots \boldsymbol{\theta}_{N_\text{pop}})$ to \\
	 \hspace*{6.5mm} obtain responses $\objSet_{k,1}^\prime, \dots, \objSet_{k,M}^\prime,\mathcal{L}_k^\prime$ \\[3pt]
	8: \quad Augment data with new evaluations: \\
 	\hspace*{6.5mm} \hbox{$\mathcal{D}_{k+1} = \left( \paramSet_k \ \tilde\cup \  \paramSet^\prime_k ,  \objSet_{k,1} \ \tilde\cup \  \objSet^\prime_{k,1},  \dots , \mathcal{L}_{k} \ \tilde\cup \  \mathcal{L}^\prime_{k}   \right)$}
 	 \\[3pt]
	9: \quad Non-Dominated Sorting of $\mathcal{D}_{k+1}$\\[3pt]
	10: \   Sort each Domination-Rank of $\mathcal{D}_{k+1}$ by Crowding \\ \hspace*{6.5mm}Distance\\[3pt]
	11: \ Truncate the elements of $\mathcal{D}_{k+1}$ to population size  $N_\mathrm{pop}$\\ \hspace*{6.5mm} based on  sorting\\[3pt]
	12: \textbf{end for} 
	\caption{NSGA-II \label{Algo:NSGAII}}	
\end{algorithm}

\subsection{Multiple Objective Particle Swarm Optimization}
MOPSO is another very popular evolutionary algorithm for multi-objective optimization. It is an extension of the well-known particle swarm optimization (PSO) for handling multi-objective optimization problems and uses a secondary repository to store the global best particles that are used for guiding the movement of particles in future iterations. 

In our studies we use the MOPSO implementation \cite{MartinezCagigal.2019} which is based on the work proposed in \cite{Coello.2004} and \cite{MargaritaReyes.2005}.
Algo.~\ref{Algo:MOPSO} outlines the procedure of the MOPSO implementation. In Step~1 and 2 an initial population $\Theta_{0}$, i.e. a set of parameterizations, with $N_\text{pop}$ particles is randomly initialized and the corresponding objective function values are obtained. Afterwards, the non-dominated solutions are determined and stored to the repository. Furthermore, the search space explored so far is subdivided into hypercubes (adaptive grid) and the particles are assigned to these hypercubes based on their objective function values.

Then the Steps 5 to 7 are processed cyclically until the maximum number of generations $N_\text{gen}$ is reached. In Step~5, a new population of particles $\Theta_{k+1}$ is obtained by first updating their positions and velocities using information from the adaptive grid, then mutation is performed, and finally the boundaries for each particle are checked. For this new population, the objective function values are queried in Step 6. Based of the results, the repository and the adaptive grid is updated in Step~7. During this process, the repository is truncated to the maximum size $N_\text{rep}$ if necessary.

\begin{algorithm}[t] 
	1: Generate an initial population $\Theta_{1} = (\boldsymbol{\theta}_{1,1}, \dots \boldsymbol{\theta}_{1,N_\text{pop}})$\\[3pt]
    2: Query objective function with $\Theta_{1}$ to obtain responses\\ 
       \hspace*{3.2mm} $(\objSet_{1,1}, \dots, \objSet_{1,M},\mathcal{L}_1)$ \\[3pt]
    3: Add non-dominated solutions to repository and generate\\
       \hspace*{3.2mm} adaptive grid\\[3pt]
	4: \textbf{for} k = 1, 2, \dots $N_\text{gen}$ \textbf{do} \\[3pt]
    5: \quad Update speeds and positions, perform mutation and\\
       \hspace*{6.5mm} check boundaries to get new population\\
       \hspace*{6.5mm} $\Theta_{k+1} = (\boldsymbol{\theta}_{k+1,1}, \dots \boldsymbol{\theta}_{k+1,N_\text{pop}})$\\[3pt]
	6: \quad Query objective function with $\Theta_{k+1}$ to obtain responses\\
	 \hspace*{6.5mm} $(\objSet_{k+1,1}, \dots, \objSet_{k+1,M},\mathcal{L}_{k+1})$ \\[3pt]
	7: \quad Update repository and adaptive grid\\[3pt]
	8: \textbf{end for} 
	\caption{MOPSO \label{Algo:MOPSO}}	
\end{algorithm}

In order to handle the crash constraints, we use a modified version of the constraint handling original proposed in \cite{Coello.2004}. The original version checks each time two particles are compared whether their constraints are satisfied. If both are feasible ($l(\params) = 1$), the dominating particle is the winner. If one is feasible and the other is infeasible, the feasible wins. If both are infeasible, then the particle with the lowest amount of constraint violation is used. The latter is not possible in our case, since the function $l(\params)$ does not provide any information about the amount of constraint violation. Therefore, we randomly select the winning particle. All other cases are treated in the same way as the original proposed constraint handling.

For our studies, we chose $N_\text{pop} = 100$ as the number of particles, $N_\text{rep} = 250$ as the repository size, and $N_\text{gen} = 50$ as the maximum number of generations.

\section{Results} \label{sec:Results}

\subsection{Evaluated Optimizers}

In total 5 different BO variants are evaluated:

\begin{itemize}
	\item TSEMO-1-C: MOBO with TSEMO as the acqusation function, constant batch size of one and without VDP.
	\item TSEMO-A-C: MOBO with TSEMO as the acqusation function, variable batch size and without VDP.
	\item TSEMO-1-VDP: MOBO with TSEMO as the acqusation function, a constant batch size of one and with VDP.
	\item TSEMO-A-VDP: MOBO with TSEMO as the acqusation function, variable batch size and with VDP.
	\item EIM-1-VDP: MOBO with EIM as the acqusation function, a constant batch size of one and with VDP.
\end{itemize}

If VDP (cf. Sec. \ref{sec:VDP}) is not used, constant objective values are assigned to the failed evaluations. The BO variants are compared with NSGA-II and MOPSO. As an additional benchmark, random sampling (Rand) is performed by drawing parameterizations from a uniform distribution bounded by the box constraints. As the last benchmark, grid search (Grid) is used. For grid search, each parameter $\theta_i$ is discretized in 6 equally spaced levels. Based on this discretization, the parameter space is evaluated full factorial.

\subsection{Metrics} \label{sec:ResultsMetrics}

As the main performance metric, the hypervolume (HV) indicator (e.g. \cite{10.1145/3453474                 }) is used. It is the most common metric for comparing multi objective optimization algorithms \cite{7360024}.  
In order to minimize the impact of initial sample quality on optimizer performance, each of the algorithms is run with ten different seeds. From these runs, the median HV indicator is calculated and used to compare the different optimization approaches. In addition to comparing the median, a hypothesis test is used to evaluate the statistical significance of the differences. It cannot be assumed that the distribution of the HV indicator follows a Gaussian. Therefore, the non-parametric one sided Wilcoxon rank sum test is used with a significance level of \SI{5}{\%}.

On average, one evaluation accounts for approximately \num{8.4e3} simulation times steps and takes around 20 seconds. The objective function evaluation time differs depending on the parameterization of the algorithm. Therefore, simulation steps are used as a measure of computational effort instead of objective function evaluations.   
In contrast to the evaluated metaheuristics, BO has a large overhead caused by Steps 4-7 in Algo. \ref{Algo:BOMultObjCrashConstrnt}. Therefore, in order to evaluate the practical usability of the optimizers, it is not sufficient, to only compare the computing effort of the objective function evaluations. Instead, the overhead is converted to equivalent simulation time steps by dividing the overhead time by the average time required for one simulation time step. The converted overhead is added to the simulation steps, to get the total computing effort in terms of simulation time steps.

\subsection{Comparison of BO Variants}
\begin{figure} [th]
    \centering
    \includegraphics[page=1,width=1.0\linewidth]{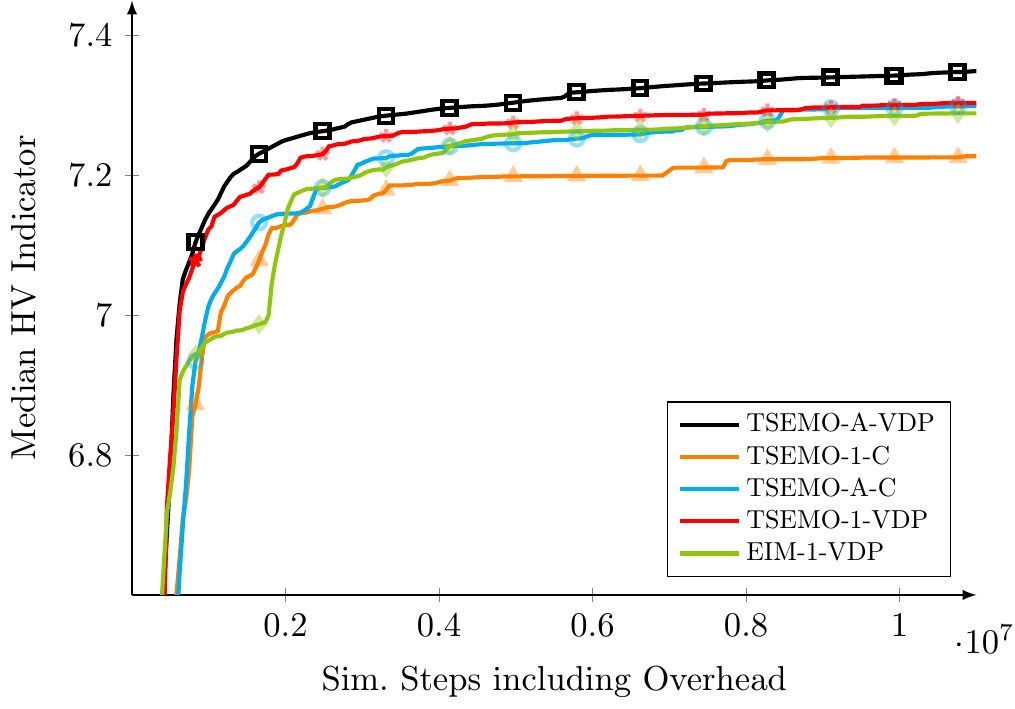}
    \caption{HV indicator as a function of simulation steps including algorithmic overhead. Markers in light colors indicate that the HV indicator of the respective algorithm is statistically significantly worse than the best BO variant.}
    \label{fig:sampleEfficiencyShortBO}
\end{figure}

Fig. \ref{fig:sampleEfficiencyShortBO} shows the progression of the HV indicator for the evaluated BO variants and a budget of $10^7$ steps $(\sim 1180 \ \mathrm{evals})$. It can be observed, that TSEMO with flexible batch size and virtual data points to handle crash constraints (TSEMO-A-VDP) performs best. If a fixed budget of one evaluation per iteration is used (TSEMO-1-VDP), a decrease in performance can be observed with increasing number of steps. After around \num{1.7e6} steps $(\sim 200 \ \mathrm{evals})$ the difference is statistically significant. The same trend can be observed when comparing the BO variants TSEMO-A-C and TSEMO-1-C, which do not use VDP. 

\begin{figure} [h]
    \centering
    \includegraphics[page=2,width=1.0\linewidth]{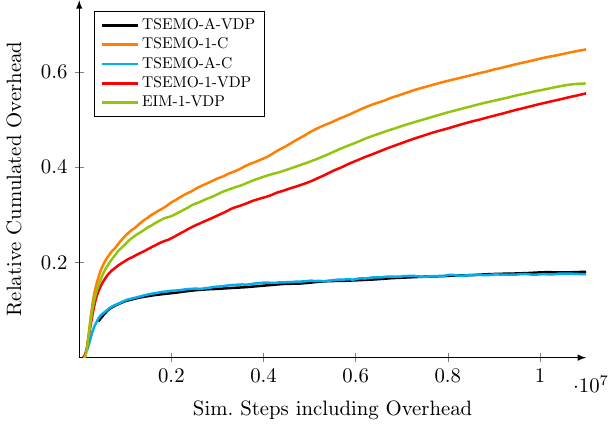}
    \caption{Relative cumulative overhead (c.f. Sec. \ref{sec:ResultsMetrics}) of the BO variants. The overhead of Rand, NSGA-II and MOPSO are not shown because they are very small $< 0.01$. }
    \label{fig:overheadShort}
\end{figure}

As an explanation, Fig. \ref{fig:overheadShort} shows the overhead of the BO variants. It can be seen that the relative cumulative overhead is bounded to under $0.2$ by adapting the batch size. Therefore in this case, trading overhead for a potential reduction in sample quality is shown to be beneficial.   

Additionally, using the adaptive VDP approach to deal with crashed simulations is shown to be significantly better than assigning a fixed objective function value. This can be observed for the cases with adaptive batch size and also for a batch size of one. In comparison to TSEMO (TSEMO-1-VDP), the expected improvement matrix criterion (EIM-1-VDP) performs worse for the application at hand.

\subsection{Overall Comparison}

\begin{figure} [H]
    \centering
    \includegraphics[page=3,width=1.0\linewidth]{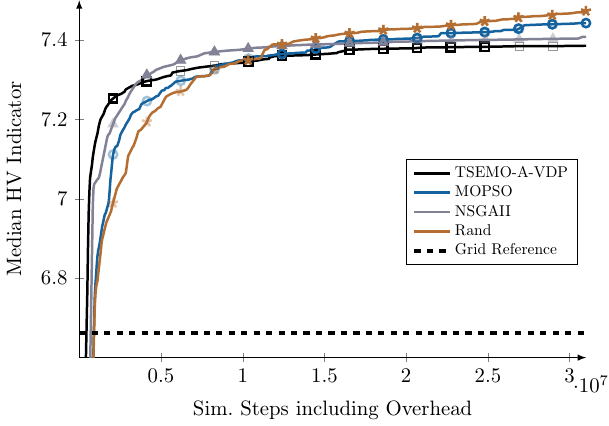}
    \caption{HV indicator as a function of simulation steps including algorithmic overhead. Markers in light colors indicate that the HV indicator of the respective algorithm is statistically significantly worse than the best performing optimizer.}
    \label{fig:BO_FinalQuality}
\end{figure}

Fig. \ref{fig:BO_FinalQuality} compares the best performing BO variant with random sampling, grid search, NSGA-II, and MOPSO for a budget of \num{3.25e7} steps $(\sim 3850 \ \mathrm{evals})$. For a small budget, i.e. until around \num{2.5e6} steps $(\sim 294 \ \mathrm{evals})$ 
, TSEMO-A-VDP performs statistically significantly better than the other optimizers.  
 For a medium budget of around \num{8.0e6} steps $(\sim 950 \ \mathrm{evals})$, NSGA-II is statistically significantly best. 
After around \num{1.2e7} steps $(\sim 1430 \ \mathrm{evals} )$ the median of random sampling (Rand) is best. MOPSO performs similarly to Rand with slight advantages for a small to medium sized budget. However, in contrast to TSEMO-A-VDP and NSGA-II, MOPSO is not statistically significantly worse than Rand at maximum budget. Importantly, TSEMO-A-VDP performs worse at maximum budget, even if overhead is not taken into account\footnote{The corresponding plot is not shown here for briefness.}. This indicates that in addition to an increase in overhead, TSEMO may suffer a decrease in sample quality if a lot of data is already acquired.

 All optimizers as well as random sampling quickly outperform grid search although grid search has a substantially higher computational effort at \num{5.24e7} steps.
This may indicate a low intrinsic dimensionality of the problem.

\subsection{Practical Implications}

\begin{figure*}
    \centering
   \includegraphics[page=4,width=1.0\linewidth]{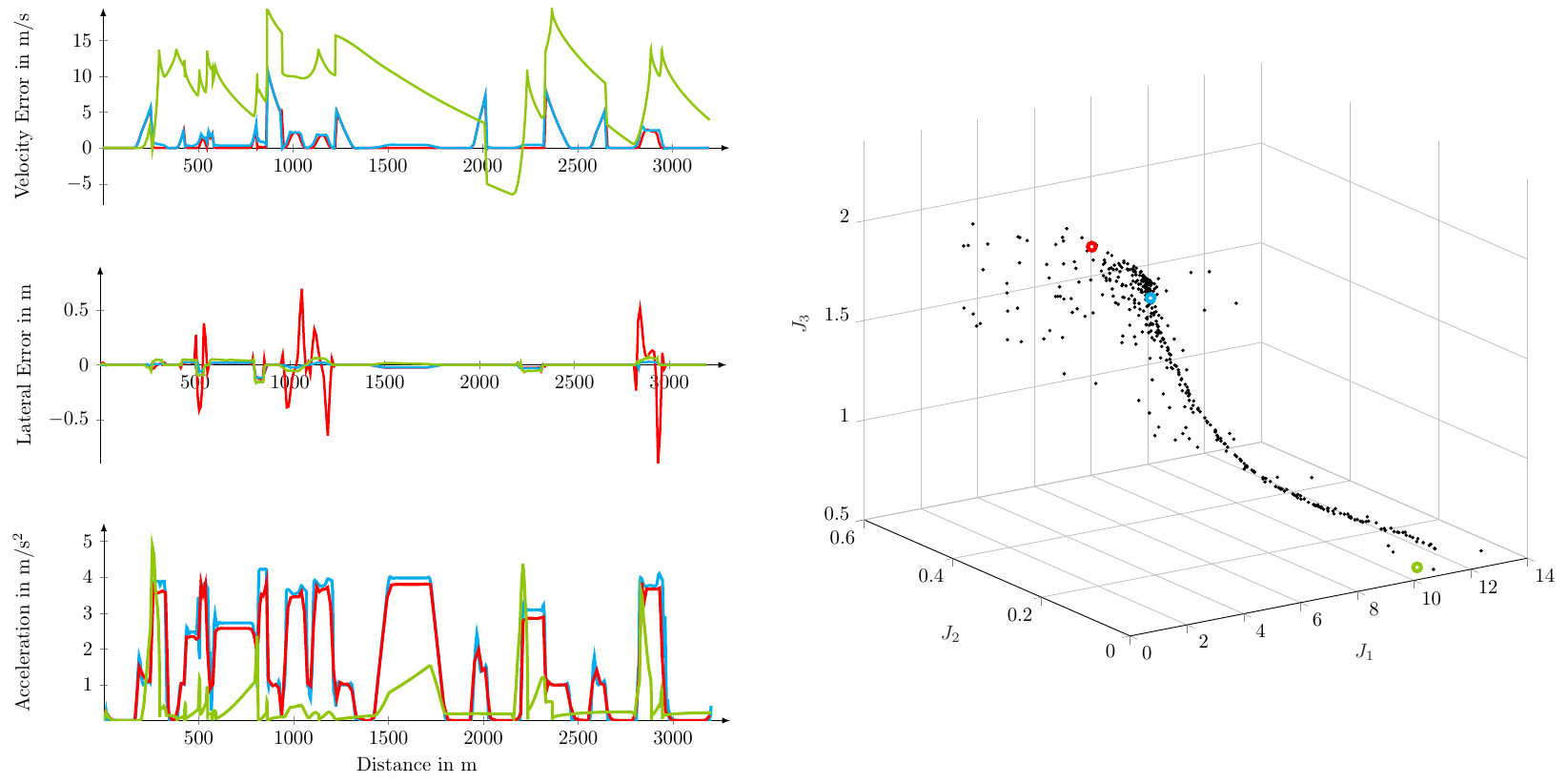}
    \caption{\textbf{Left}: Time domain plot for three different Pareto-optimal MPC parameterizations. \textbf{Right}: Corresponding objective Pareto front. The green, blue and red parameterizations focus on minimizing acceleration, lateral error and velocity error respectively. All other parameterizations located on the Pareto front (black dots) represent alternative non-dominated compromises.     }
    \label{fig:TimePlot}
\end{figure*}

Fig. \ref{fig:TimePlot} exemplary shows the time domain behavior of three different parameterizations belonging to the Pareto front. It can be observed that vastly different control behavior can be achieved with the different parameterizations. 

From a practical point of view, the knowledge of the Pareto front with its corresponding parameterizations has many advantages.
First, it enables intuitive parameterization of the vehicle guidance system: an application engineer can move on the Pareto front if he/she feels that the driving experience is too smooth or too rough. Users can also look up the corresponding optimal parameterizations without the need of deeply understanding the MPC itself or manually tune the parameters through numerous driving tests. In addition, the dimensionality for the application of vehicle guidance is significantly reduced. In the case considered here from five MPC parameters to three intuitive objectives. In other applications this ratio can vary arbitrarily. This can significantly reduces the time required for the application.

Second, the optimization may find parameterizations that even an experienced engineer would not have tried: this may increase the control performance compared to manual tuning.

Third, knowledge of the Pareto front makes it very easy to implement automatic switching of MPC parameterization: this allows, for example, a situation-dependent switching of the parameterization based on the objectives that are important for a situation. For instance, one could use a different parameterization on the highway (comfort is important) than when parking (tracking is important).

\section{Conclusions} \label{sec:conclusion}
 
 In this paper, we considered model predictive control (MPC) for the lateral and longitudinal control of autonomous vehicles and compared the sample-efficiency and the computational overhead of five different multi-objective Bayesian optmization (MOBO) versions and two metaheuristics, Multiple Objective Particle Swarm Optimization (MOPSO) and Non-dominated Sorting Genetic Algorithm II (NSGA-II).
It is shown that multi-objective MPC tuning is capable of automatically finding parameters which produce versatile closed-loop behaviors.
 The presented method essentially has two 
 advantageous features. First, the problem of applying  vehicle guidance is reduced from five dimensions in parameter space to a two-dimensional manifold in three-dimensional objective function space. Second,  the individual parameterizations represented by each point of this manifold can be interpreted very well in their resulting physical effects. How strong the dimension reduction and the interpretability is pronounced is determined by the respective application or the respective design. 
 
 From an optimization point of view, it was shown for this specific application that addressing crash constraints properly is essential to BO performance. Additionally, the overhead was bounded by adaptively choosing the sample size, increasing overall optimization speed. In comparison to other optimizers, BO was only best for a small number of objective function evaluations. For medium budgets NSGA-II is best and for large budgets, only MOPSO was not statistically significantly worse than random search. Grid search was clearly outperformed. Future work should addressed whether these findings carry over to other applications.

\section*{Acknowledgement}
This work was partially supported by the German Federal Ministry for Economic Affairs and Climate Action (BMWK) under the grant 50NA1912.

\bibliographystyle{IEEEtran}
\bibliography{lit}

\end{document}